\documentclass[journal]{IEEEtran}

\usepackage{graphicx}
\usepackage{epsf}
\usepackage{mathtools}
\usepackage{amssymb}
\usepackage{amsmath, amsfonts}
\usepackage{latexsym}
\usepackage{multirow}
\usepackage{multicol}
\usepackage[table]{xcolor}
\usepackage{color}

\usepackage{tabularx}
\usepackage{lipsum}

\usepackage{algorithm}
\usepackage{algorithmic}

\usepackage{mathrsfs}

\usepackage{fixltx2e}

\usepackage[mathscr]{euscript}

\usepackage{commath}
\usepackage{MnSymbol}

\usepackage{subcaption}

\usepackage{mathtools} 
\DeclarePairedDelimiterX\setc[2]{[}{]}{\,#1 \;\delimsize\vert\; #2\,}
\DeclarePairedDelimiterX\parth[2]{(}{)}{\,#1 \;\delimsize\vert\; #2\,}

\PassOptionsToPackage{hyphens}{url}
\usepackage[unicode=true, bookmarks=true, bookmarksnumbered=true, bookmarksopen=true, bookmarksopenlevel=1, breaklinks=false, pdfborder={0 0 0}, pdfborderstyle={}, backref=false, colorlinks=false]{hyperref}

\usepackage{theorem}
{
\theorembodyfont{\rmfamily}
\newtheorem{assumption}{Assumption}

\newtheorem{lemma}{Lemma}

\newtheorem{theorem}{Theorem}
}

\definecolor{orange}{RGB}{255,127,0}
\definecolor{blue}{RGB}{0,0,255}
\definecolor{red}{RGB}{255,0,0}
\definecolor{green}{RGB}{50,160,50}
\definecolor{grey}{RGB}{125,120,125}
\definecolor{purple}{RGB}{125,0,125}

\begin{document}
{
\title{{\fontsize{15}{2}\selectfont Spatiotemporal Analysis of Shared Situation Awareness among Connected Vehicles}}

\author
{
Seungmo Kim, \textit{Senior Member}, \textit{IEEE}

\vspace{-0.3 in}

\thanks{S. Kim, who can be reached at seungmokim@georgiasouthern.edu, is with the Department of Electrical and Computer Engineering, Georgia Southern University in Statesboro, GA, USA.

This work was supported by the National Science Foundation (NSF) via award \#2138446.}
}

\maketitle
\begin{abstract}
Shared situation awareness (SSA) has been garnering explosive interest in various applications for intelligent transportation systems (ITS). In addition, the delay-constrained nature of supporting vehicular networks makes it critical to precisely analyze the performance of a SSA procedure. Extending the relevant literature, this paper provides an analysis framework that evaluates the performance of SSA in spatial and temporal aspects simultaneously. Specifically, this paper provides a closed-form probability distribution for the length of time taken for constitution of a SSA among a group of connected vehicles. This paper extends the calculation to investigation of feasibility of SSA in supporting various types of safety messages defined by the SAE J2735.
\end{abstract}


%%%%%%%%%%%%%%%%%%%%%%%%%%%%%%%%%%%%%%%%%%%%%%%%%%%%%%%%%%%%%%%%%%%%%%%%%%%%%%%%%%%%%%%%%%%%%%%%%%%%%%%%%%%%%%%%%%%%
\section{Introduction}\label{sec_intro}
\textit{How fast can a group of connected vehicles reach a shared situation awareness?} This paper targets to find an answer to the question. Specifically, it aims to extend the state of the art that will be laid out in Section \ref{sec_related}, via the following unique contributions:
\begin{itemize}
\item Formulates an accurate mathematical framework that quantifies the \textit{latency for a shared situation awareness (SSA)} among a group of connected vehicles
\item Investigates the feasibility of SSA in supporting \textit{practical ITS scenarios}
\end{itemize}

%\vspace{0.1 in}

\textit{Situation awareness (SA)} is defined as perception of environmental elements and events in relation to time or space \cite{End11}. According to the given definition, \textit{SSA} is characterized as ``a state in which all members of a group share the same SA'' \cite{End11}. This shared awareness enables individuals, even those in isolated situations, to receive support from nearby peers. The collective awareness fostered by others contributes to enhancing the overall understanding within the entire group.

Recently, there have been proposals on applying SSA as part of intelligent transportation system (ITS) \cite{tits11}. As one of key technologies supporting the SSA, vehicle-to-everything (V2X) communications have been taking the central role in enabling vehicles to exchange safety-critical data \cite{fcc20164}. Not only existing standards such as dedicated short-range communications (DSRC) \cite{dietrich_gc18}-\cite{11pstd}, emerging technologies such as cellular V2X (C-V2X) \cite{misener}-\cite{sunuwar_secon24} also burgeon as key drivers of V2X networking.

Nevertheless, several critical unseen problems have been encountered as it is only a recent but very explosive change that vehicles communicate to achieve a SSA. First, V2X communications can be competitive especially when a large number of vehicles attempt to access \cite{misener}. Second, it is not a trivial task to formulate the characteristics of a SSA due to (i) necessity to process a large amount of data and (ii) dynamicity and complicacy of environment with too many parameters \cite{ssa_complicacy}. Third, while SSA is delay-constrained, it is a challenge to write a formal analysis framework for the delay because the temporal aspect of a V2X network is closely connected to the relative positions and moving directions of vehicles \cite{bennis}.

The intricate nature of the situation underscores the urgent necessity for a detailed examination of \textit{spatiotemporal} aspects. In response, this paper aims to develop an analytical framework focusing on the performance modeling of a SSA within a network of connected vehicles. More specifically, the paper engages in a stochastic analysis to pinpoint a probability distribution that quantifies the length of time taken for completion of a SSA.

\section{Related Work}\label{sec_related}
The recent literature on Intelligent Transportation Systems (ITS) has prominently featured SSA, particularly in the context of applications for connected vehicles. Previous studies have concentrated on the interaction between humans and vehicles, exploring areas such as the design of human-vehicle interfaces for autonomous driving \cite{WaS19} and evaluating potential threats during braking scenarios \cite{HwC19}.

There is also a body of existing work taking a systematic point of view on the process of achieving a SSA. A group decision making (GDM) problem was solved based on a fuzzy ontology theory for consensus constitution for a multi-unmanned-vehicle system \cite{collective}. The literature evolved in the direction of enabling a group of connected vehicles to execute cooperative perception without any assumption of central authority and specific protocol \cite{neuron}.

Further application of the SSA to an ITS has also appeared in existing work from pedestrian localization at a roadside unit (RSU) \cite{OjV20} to enhancement of key functionalities for smart cities such as security and scalability \cite{blockchain}.

While the literature stretched the discussion toward the line of diverse ITS-relevant applications, it showed relatively less effort on investigating how exactly a SSA is achieved in a dynamic environment such as a V2X network. Earlier studies \cite{haenggi_tits16}\cite{vanet_access19} has established a probability model for the performance of a V2X network based on the stochastic geometry. However, the ``spatial'' aspect of the model was one-dimensional, which was too simplistic to reflect impacts of vehicles' positions and movements in a higher-dimensional space. Also, while existing analysis \cite{haenggi10} presents the distribution of distance in a point process, this paper uniquely extends the existing analysis to the distribution of the ``time length'' for completion of knowledge propagation. This is where both space and time aspects of a SSA constitution process are considered in a single analysis framework. As such, this paper puts particular emphasis on analyzing the specifics of the process itself that a SSA is constructed among a group of connected vehicles.

\begin{figure}
\centering
\includegraphics[width=\linewidth]{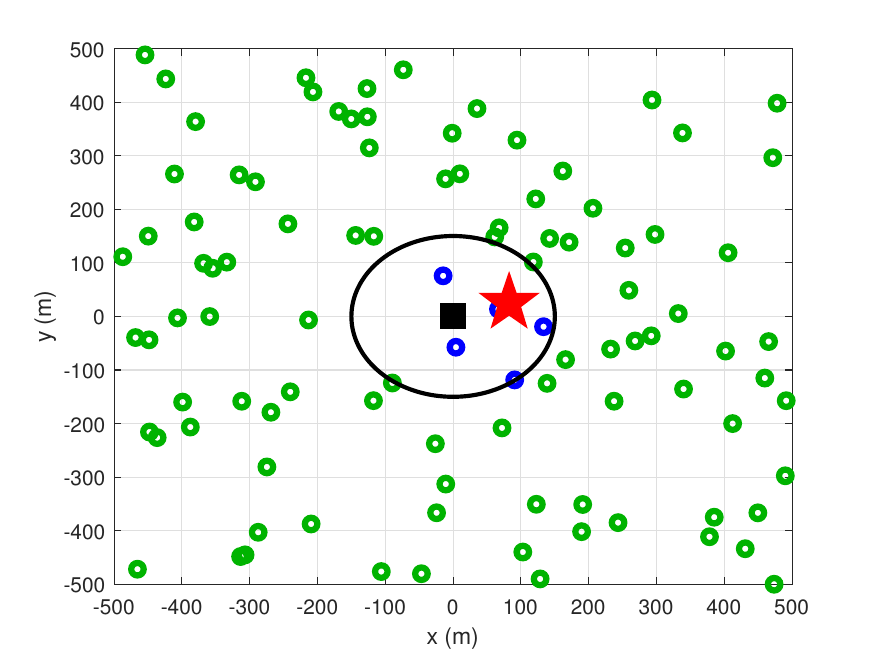}
\caption{A single snapshot of the mobile V2X system for scenario one with $\lambda$ = 100/$\left|\mathbb{A}^2\right|$ and $\left(W,D\right)$ = (1000 m, 1000 m) (The black square at the origin is the tagged vehicle, and the black circle around it indicates the transmission range of the tagged vehicle. A blue circle indicates a vehicle neighboring to the tagged vehicle, while a green circle gives a vehicle that is outside of the tagged vehicle's transmission range. The red star inside the circle is the target of awareness by the tagged vehicle.)}
\label{fig_model}
\end{figure}

%%%%%%%%%%%%%%%%%%%%%%%%%%%%%%%%%%%%%%%%%%%%%%%%%%%%%%%%%%%%%%%%%%%%%%%%%%%%%%%%%%%%%%%%%%%%%%%%%%%%%%%%%%%%%%%%%%%%
\section{System Model}\label{sec_model}
\subsection{Geometry}\label{sec_model_geometry}
Fig. \ref{fig_model} shows a snapshot for an exemplary distribution of 100 vehicles in a two-dimensional space (which will be denoted by $\mathbb{A}^2$), whose width and depth are $W$ m and $D$ m, respectively. This model is designed to simulate a small, dense area with numerous vehicles exchanging their observations with each other. The model generates vehicles (with a uniform distribution) in a restricted area and places a set of vehicles randomly (with another independently and identically distributed (i.i.d.) uniform distribution) within the defined space of $\mathbb{A}^2$. The position of the $i$th vehicle is denoted by $\mathtt{x}_{i}=\left(x_{i},y_{i}\right) \in \mathbb{A}^2$. 

Elaborating on the nodes distribution, we assume a Poisson point process (PPP) with an intensity $\lambda$, wherein the number of vehicles distributed within $\mathbb{A}^2$ follows a Poisson($\lambda$) random variable. It is important to note that based on the modeling with PPP, the uniformity property of a homogeneous point process can be held. That is, if a homogeneous point process is defined on a real linear space, then it has the characteristic that the positions of these occurrences on the real line are uniformly distributed \cite{pinsky}. Note also that the PPP is a \textit{stationary point process} where the density $\lambda$ remains constant across $\mathbb{A}^2$.

From Fig. \ref{fig_model}, notice of a \textit{generalized} environment, which is distinguished from the system models shown in other relevant work. We stress that this system model can act as the most generic form that can accommodate vehicles' movement in any direction, which, in turn, enables the system to be applied to various scenarios including flight of unmanned aerial vehicles (UAVs), lane changing, intersection, and pedestrian walking.

For generation of results that will be presented in Section \ref{sec_results}, the layout shown in Fig. \ref{fig_model} is repeated with a sufficiently large number of trials to approximate all possible positions of vehicles in the defined two-dimensional space $\mathbb{A}^2$.

Notice that we assume of a \textit{closed system} where vehicles cannot leave the $\mathbb{A}^2$ shown in Fig. \ref{fig_model}. Upon arrival at the limit of $\mathbb{A}^2$, a vehicle bounces with a reflected angle along the edge. This is to ensure that the system is assessed with a certain value of $\lambda$ kept consistent during a round of evaluation.

\begin{figure}
\centering
\begin{subfigure}{.495\textwidth}
\centering
\includegraphics[width=\linewidth]{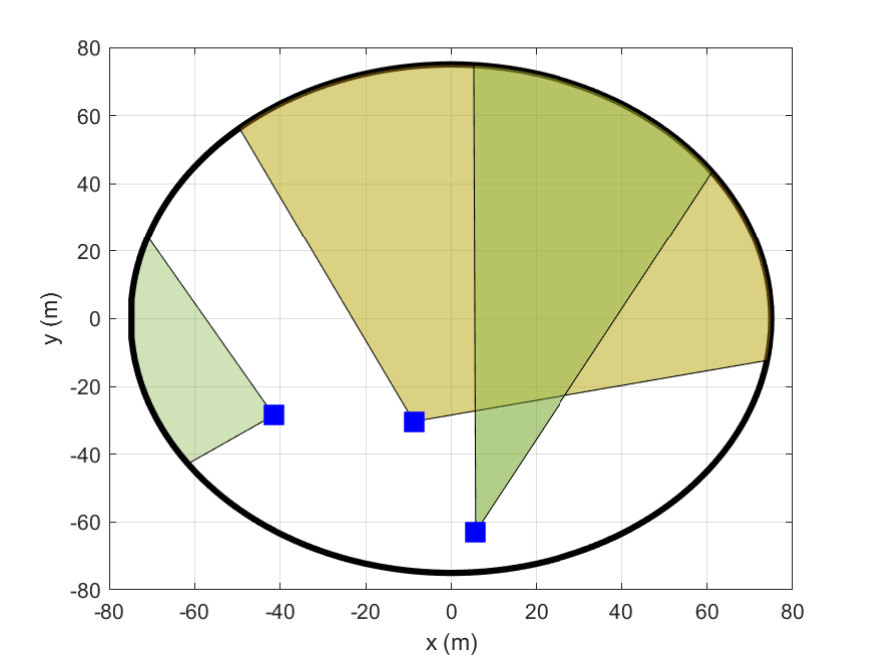}
\caption{Knowledge of each vehicle}
\label{fig_intersection}
\end{subfigure}
\hfill
\begin{subfigure}{.495\textwidth}
\centering
\includegraphics[width=\linewidth]{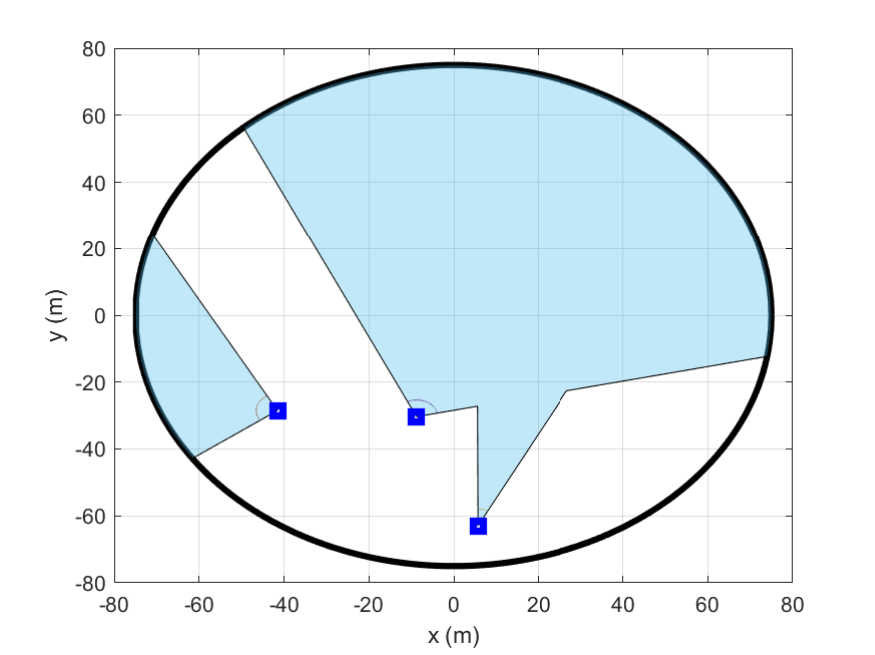}
\caption{SSA among the vehicles as a union of the knowledge}
\label{fig_union}
\end{subfigure}
\caption{Union and intersection of shared knowledge among multiple nodes (An example with the detection radius of 75 m by each vehicle.)}
\label{fig_vision}
\end{figure}

\subsection{Networking among Vehicles}\label{sec_model_network}
In the vehicular network model presented in this paper, each vehicle is assumed to be mobile. Also, an entire network is \textit{fully connected}: every node is supposed to be equipped with communication functionality and hence is able to broadcast it whenever needed. It is also significant to note that the connection type of a network is completely \textit{distributed}, \textit{e.g.}, mode 4 of C-V2X, in which no central coordinator node nor infrastructure exists. On the networking setting, we make a significant assumption on the process of constructing a SSA:

\vspace{-0.1 in}

\begin{assumption}\label{assumption_1st}
(A vehicle's selection of neighbor for SSA). \textit{A vehicle $i$ is assumed to choose ``the 1st neighbor'' (defined as the vehicle with the shortest distance within the vehicle $i$'s transmission range) for handing its SA in the process of collecting a SSA.}
\end{assumption}

We claim that this system model can generally suit currently operating V2X systems in practice including IEEE 802.11-based system such as 802.11p (also known as DSRC) \cite{11pstd} and 802.11bd \cite{80211bd}. The model can also be applied to C-V2X mode 4 where the nodes are connected directly in a distributed manner without going through the network core, e.g., sidelink-based broadcast or groupcast as defined in the 3rd Generation Partnership Project (3GPP) Release 16 \cite{tr22886}. For exchange of information among vehicles, these V2X techniques are usually based on listen-before-talk (LBT) and periodic broadcast of basic safety messages (BSMs) with no feedback available. We suggest that these methods of data exchange will suffice for propagation of a SSA fragment all over a network. The rationale is two-fold: (i) a SSA usually requires a very short delay and (ii) such ``light'' data exchange methods, albeit the possibility of sacrificing data rate, have been proved to be able to support the low-latency applications \cite{access20}.

%%%%%%%%%%%%%%%%%%%%%%%%%%%%%%%%%%%%%%%%%%%%%%%%%%%%%%%%%%%%%%%%%%%%%%%%%%%%%%%%%%%%%%%%%%%%%%%%%%%%%%%%%%%%%%%%%%%%
\section{Analyses on SSA Completion Time}\label{sec_analysis}
We are interested in analyzing the performance of SSA achieved among a group of connected vehicles. Given the fact that we are deploying SSA in a connected-vehicle network, we put particular focus of the analysis on measuring the length of time until achievement of a SSA. The rationalizations for the two fronts are: (i) a V2X network is delay-constrained and (ii) the vehicles constituting a network are supposed to be mobile and hence the coverage depends on the mobility, respectively.

\subsection{Cooperative SSA}\label{sec_analysis_ssa}
At each vehicle, we assume a \textit{limited sight}: each node has a certain finite value for the \textit{width} and \textit{length} of its sight. This means that each node's sight forms an \textit{arc's shape} in front of it. We model the width and length of each arc as a random variable (i.e., uniformly distributed), considering that the sight varies according to numerous factors such as the driver's ability, existence of physical obstacles, etc.

Fig. \ref{fig_vision} illustrates an example of this scenario with three vehicles participating in observation of a target area. (Notice that the big black circle and the small blue squares indicate a target area and vehicles aiming to observe the target area, respectively.)

Given that a full SSA is achieved when 100\% of the circle's area is covered, one can say from Fig. \ref{fig_union} that the three vehicles are achieving a SSA at the rate of
\begin{align}\label{eq_R_ssa}
\mathsf{R}_{\text{ssa}}\left(t_{0}\right) = \displaystyle \Big( \bigcup_{t \le t_{0}} \bigcup_{i \in \mathcal{S}_{\text{ssa}}} \left|\mathbb{A}_{i,t}^2\right| \Big) \cdot \Big( \left|\mathbb{A}_{\text{tgt}}\right| \Big)^{-1}
\end{align}
where $\left|\mathbb{A}_{i,t}^2\right| \le \left|\mathbb{A}_{\text{ssa}}^2\right|$. Notice of the parameters as follows: $\mathcal{S}_{\text{ssa}}$ gives a set of vehicles contributing the SSA on a target area; $\mathbb{A}_{i,t}^2$ gives the space formed by the vision of vehicle $i$ at time instant $t$; and $\mathbb{A}_{\text{tgt}}^2$ gives the space defined by a target area. It is significant to notice from Eq. (\ref{eq_R_ssa}) that the rate $\mathsf{R}_{\text{ssa}}$ varies according to not only vehicles but time, reflecting mobility of the vehicles. That is, each vehicle contributes with different angles of vision as it moves within the target area. This depicts that every part of the targeted area is seen by at least one of the users and a communication link is assumed between them to share statuses for the whole area.

We model the cooperative SSA as the \textit{subset sum problem}, which is defined as a decision problem to decide whether any subset of the integers sum to precisely a target sum $T$. The problem is known to be \textit{NP-complete} \cite{arxiv18}, which makes this problem particularly challenging and thus valuable to solve.

\subsection{Latency for SSA}\label{sec_analysis_latency}
By $h$, let us denote a ``hop'' between two neighboring vehicles, which can be formally written as
\begin{align}\label{eq_hop}
h = d/v \stackrel{(a)}{=} d/c
\end{align}
where $d$ denotes the distance of a hop between two neighboring nodes and $v$ gives the velocity of propagation of a V2X communications signal. The equivalence (a) is from that propagation of a microwave signal over the air takes the speed of light, which is written as $c$.

Now, we stress that $d$ is a value of a random variable since $\mathtt{x}_{i}$ is, as was explained in Section \ref{sec_model_geometry}. Let $D$ and $H$ denote the random variables for $d$ and $h$, respectively. Also note that the distance to the $n$th neighbor from a reference point in a PPP---denoted by $d$ in Eq. (\ref{eq_hop})---was found to follow a generalized Gamma distribution \cite{haenggi10}. With the density denoted by $\lambda$ as was mentioned in Section \ref{sec_model_geometry}, let us denote a single-hop distance by a random variable $D \sim \text{Gamma}\left(k, \lambda\right)$ where $k$ denotes the shape parameter of the probability density function (PDF).

We know that the number of vehicles distributed within $\mathbb{A}^2$ follows a Poisson distribution, as was explained in Section \ref{sec_model_geometry}. Now, we notice that the knowledge propagation in a SSA process takes multiple hops among the participating vehicles. Since each vehicle can only hand a message to vehicles within its transmission range (which we call ``neighboring vehicles'' from this point), a detailed analysis has to be done within a vehicle's transmission range, for further advancement in the analysis.

\begin{figure}
\centering
\includegraphics[width=\linewidth]{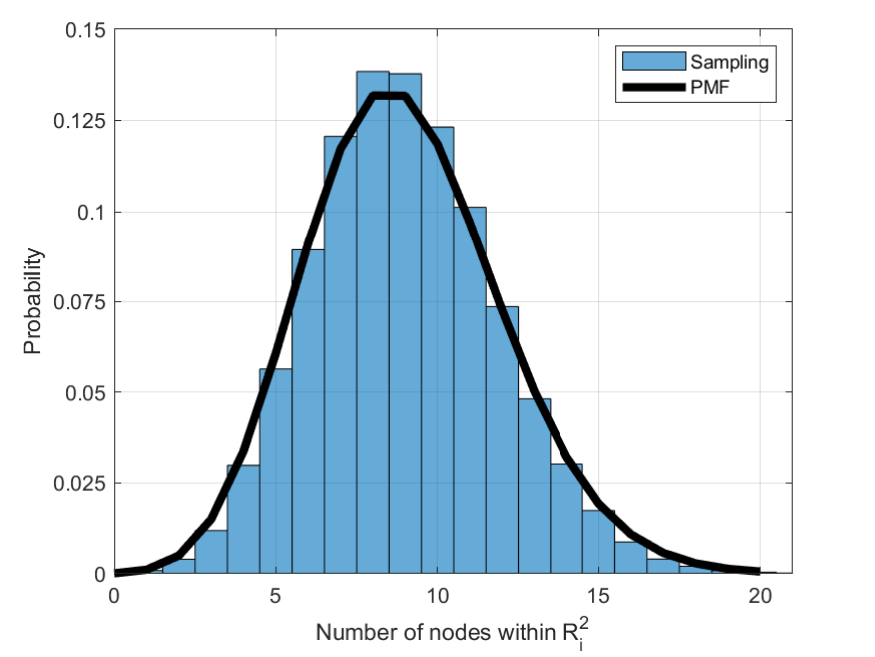}
\caption{Comparison of sampling and PDF of the Poisson($\rho\lambda$) for proof of Lemma \ref{lemma_poisson_subset}}
\label{fig_poisson_subset}
\end{figure}

\begin{lemma}\label{lemma_poisson_subset}
(Distribution of the number of vehicles within a subset of $\mathbb{A}^2$) \textit{Let the transmission range of an arbitrary vehicle $i$ be denoted by $\mathbb{A}_{i}^2, \hspace{0.02 in} \forall i \in \mathbb{N}$, which, as such, yields $\mathbb{A}_{i}^2 \in \mathbb{A}^2$. Given that the number of vehicles counted in $\mathbb{A}^2$ follows Poisson($\lambda$), that counted in $\mathbb{A}_{i}^2$ follows Poisson($\rho\lambda$) where $\rho = \left|\mathbb{A}_{i}^2\right| / \left|\mathbb{A}^2\right|$.}
\end{lemma}

\vspace{-0.07 in}

\textit{Proof:}
A relevant proof for Lemma \ref{lemma_poisson_subset} can be found in Eq. (5.27) of \cite{pinsky}. In addition, as a means to strengthen the proof, we conducted a comparison between (i) sampling and (ii) probability mass function (PMF) of the Poisson($\rho\lambda$) as proposed in the proposition. Notice that the PMF is given by $\mathbb{P}\left[N = n\right] = \left(\rho\lambda\right)^{n}\exp\left(-\rho\lambda\right) / k!$. Fig. \ref{fig_poisson_subset} shows the comparison where the sampling and the PMF match with a substantial deal of accuracy, which reinforces the proof.
\hfill$\blacksquare$

\vspace{0.12 in}

The distribution of $D$ has been proposed as a generalized Gamma \cite{haenggi10}. We extend the existing work to the direction of modeling a $n$th hop in a multi-hop path. As presented in Assumption \ref{assumption_1st}, we only focus on the 1st neighbors in each vehicle participating in a SSA. Now, revisiting Eq. (\ref{eq_hop}), we can proceed to finding the distribution of $h = d/c$. It has been known that scaling a random variable following $\text{Gamma}\left(k,\lambda\right)$ by a constant $1/c$ yields another Gamma random variable, $\text{Gamma}\left(k,c\lambda\right)$.
\begin{lemma}\label{lemma_hop_gamma}
(Distribution of a hop). \textit{Thus, the PMF of $H$ is formally written as}
\begin{align}\label{eq_hop_pdf}
\mathbb{P}\left[H = h\right] = \displaystyle \frac{\left(c\lambda\right)^{k}}{\Gamma\left(k\right)} h^{k-1} e^{-c\lambda x}
\end{align}
\end{lemma}

\textit{Proof:}
We refer interested readers to references such as \cite{textbook12} for mathematical details for Lemma \ref{lemma_hop_gamma} where it can be easily found as ``scaling property of Gamma distribution.''
\hfill$\blacksquare$

\vspace{0.2 in}

\begin{figure}
\centering
\includegraphics[width=\linewidth]{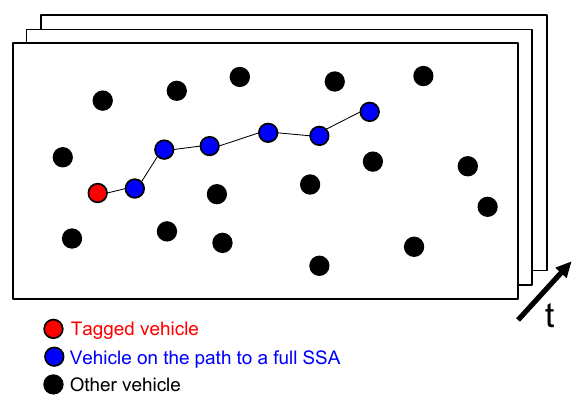}
\caption{Abstraction of constitution of a path until achievement of a SSA}
\label{fig_tree}
\end{figure}

We remind from Fig. \ref{fig_tree} that a SSA is constructed among the 1st neighbors in each vehicle's transmission range. As such, one can note that a complete path can be characterized as a sum of a certain number of hops. It is also important to note that on the path, each hop $h_{j}$ is independent and identically distributed (i.i.d.) with each other since a vehicle does not coordinate only except simply adding up what it has found. With all these considered, for a SA construction path, one can formally write the total number of hops that have been taken to achieve a SSA among $N$ vehicles as
\begin{align}\label{eq_hop_sum}
h_{\text{ssa},i} &= \displaystyle \sum_{j \in \mathcal{S}_{\text{ssa}}} h_{j}
\end{align}
where $\mathcal{S}_{\text{ssa}}$ denotes a set of vehicles participating in constitution of a SSA, and $h_{j}$ gives each hop in an arbitrary path $j$. Now, building Lemmas \ref{lemma_poisson_subset} and \ref{lemma_hop_gamma} on Eq. (\ref{eq_hop_sum}), we can finalize the analysis by finding the distribution of $h_{\text{ssa},i}$.

Assume that each vehicle has the same transmission range, which is a practical assumption since it is likely that a same technical spec on the transmit power is applied at the stage of manufacturing based on a certain standard such as DSRC \cite{11pstd}\cite{80211bd} and C-V2X \cite{tr22886}. This yields that the number of neighboring vehicles in each hop follows a same Poisson($\rho\lambda$) since $\rho$ remains the same due to $\left|\mathbb{A}_{i}^2\right|$ remaining identical at an arbitrary vehicle $i$.

\vspace{0.05 in}

\begin{theorem}\label{theorem_gamma_sum}
(Distribution of latency for a complete SSA among a group of connected vehicles). \textit{Building on Lemma \ref{lemma_hop_gamma}, the total number of hops in path $j$ taken for completing a full SSA can be characterized as a ``sum'' of $N$ Gamma random variables following $\text{Gamma}\left(k,c\lambda\right)$ whose PMF has been given in Lemma \ref{lemma_hop_gamma}. Let us denote this sum as random variable $H_{\text{ssa},i}$. We claim that $H_{\text{ssa},i} \sim \text{Gamma}\left(Nk,c\lambda\right)$ whose PDF is given by}
\begin{align}\label{eq_Hssa_PDF}
\mathbb{P}\left[H_{ssa} = h\right] = \displaystyle \frac{\left(c\lambda\right)^{Nk}}{\Gamma\left(Nk\right)} h^{Nk-1} e^{-c\lambda x}
\end{align}
\end{theorem}

\vspace{-0.05 in}

\textit{Proof:} We bring this knowledge to discussing Theorem \ref{theorem_gamma_sum} where we are attempting to find the distribution of a sum of $N$ Gamma random variables. While the general information for this proof can be found in the literature such as \cite{gammasum85}, we apply the proving method to write the distribution of $H_{\text{ssa}} = \sum_{j=1}^{N} H_{j}$ precisely. Let us start from finding the moment generating function (MGF) of $H_{j}$:
\begin{align}\label{eq_H_mgf}
M_{H}\left(t\right) &= \mathbb{E}\left[e^{tH}\right]\nonumber\\
&= \displaystyle \int_{0}^{\infty} e^{th} \frac{\lambda^{k}}{\Gamma\left(k\right)} h^{k-1} e^{-\lambda h} \text{d}h\nonumber\\
&= \displaystyle \int_{0}^{\infty} \frac{\lambda^{k}}{\Gamma\left(k\right)} h^{k-1} e^{\left(t-\lambda\right) h} \text{d}h
\end{align}
With substitution of $u = \left(\lambda - t\right)h$, which yields $\text{d}h = \left(\lambda - t\right)^{-1} \text{d}u$, Eq. (\ref{eq_H_mgf}) can be rewritten as
\begin{align}\label{eq_H_mgf_2}
M_{H}\left(t\right) &= \displaystyle \int_{0}^{\infty} \frac{\lambda^{k}}{\Gamma\left(k\right)} \left(\frac{u}{\lambda -t}\right)^{k-1} e^{u} \frac{1}{\lambda -t} \text{d}u\nonumber\\
&= \displaystyle \frac{\lambda^{k}}{\Gamma\left(k\right)\left(\lambda -t\right)^{k}} \int_{0}^{\infty} u^{k-1} e^{u} \text{d}u\nonumber\\
&= \displaystyle \frac{\lambda^{k}}{\Gamma\left(k\right)\left(\lambda -t\right)^{k}} \Gamma\left(k\right)\nonumber\\
&= \left(\frac{\lambda}{\lambda -t}\right)^{k}
\end{align}

Now, we note that the MGF of a sum of multiple random variables is the product of their MGFs. Let $M_{X}\left(s\right)$ denote the MGF for any general random variable $X$. It has already been known that for $Y = X_{1} + X_{2} + \cdots + X_{N}$, the MGF of the sum is the product of the MGFs, which is formally written as $M_{Y}\left(t\right) = \mathbb{E}\left[e^{tY}\right] = \mathbb{E}\left[e^{t\left(X_{1} + X_{2} + \cdots + X_{N}\right)}\right] = \mathbb{E}\left[e^{tX_{1}}e^{tX_{2}}\cdots e^{tX_{N}}\right] = M_{X_{1}}\left(t\right) M_{X_{2}}\left(t\right) \cdots M_{X_{N}}\left(t\right)$.

With all these considered, the MGF of $H_{\text{ssa}} = \sum_{j=1}^{N} H_{j}$ is given by
\begin{align}\label{eq_Hssa_mgf}
M_{H_{\text{ssa}}}\left(t\right) &= \prod_{j=1}^{N} M_{H_{j}}\left(t\right) = \left(\frac{\lambda}{\lambda -t}\right)^{Nk}
\end{align}
The MGF given in Eq. (\ref{eq_Hssa_mgf}) tells that $H_{\text{ssa}}$ follows a Gamma distribution with the shape parameter of $Nk$. Since the summation had no impact on the other parameter $c\lambda$ of the distribution, it finally yields that $H_{\text{ssa}} \sim \text{Gamma}\left(Nk, c\lambda\right)$, which completes the proof.
\hfill$\blacksquare$

%%%%%%%%%%%%%%%%%%%%%%%%%%%%%%%%%%%%%%%%%%%%%%%%%%%%%%%%%%%%%%%%%%%%%%%%%%%%%%%%%%%%%%%%%%%%%%%%%%%%%%%%%%%%%%%%%%%%
\section{Numerical Results}\label{sec_results}
This section articulates the computer simulations that we undertook as a means to confirm the theoretical findings that were presented in Section \ref{sec_analysis}.

We rationalize that simulation would accomplish the best efficiency as the main method to evaluate the performance of this paper's system model based on two advantages \cite{fire08}: (i) the capability of testing a wide diversity of parameters on a broad range of values within a relatively short period of time; and (ii) the ability to execute computations without being caught up with restrictions/errors caused by computing/experimental factors including hardware, compiler, language, etc.

\subsection{Parameters and Setup}\label{sec_results_parameters}
This paper focuses on examining the shared SA performance within a V2X network that operates in a \textit{distributed} manner. We argue that considering a distributed V2X network is practical, considering the predominance of DSRC [FIND NEW] and C-V2X mode 4 [A 3GPP TS], both of which are operated in a distributed fashion. Among the key parameters are a 100-msec inter-broadcast interval between two BSMs and a transmission range of \{75,100\} m.

The spatial dimensions of the two-dimensional space, denoted as $\mathbb{A}^{2}$, where the nodes are dispersed, are defined as 1,000 m by 1,000 m, as depicted in Fig. \ref{fig_model}. Additional details regarding geographical configurations can be referenced in Section \ref{sec_model}.

\begin{figure}
\centering
\includegraphics[width=\linewidth]{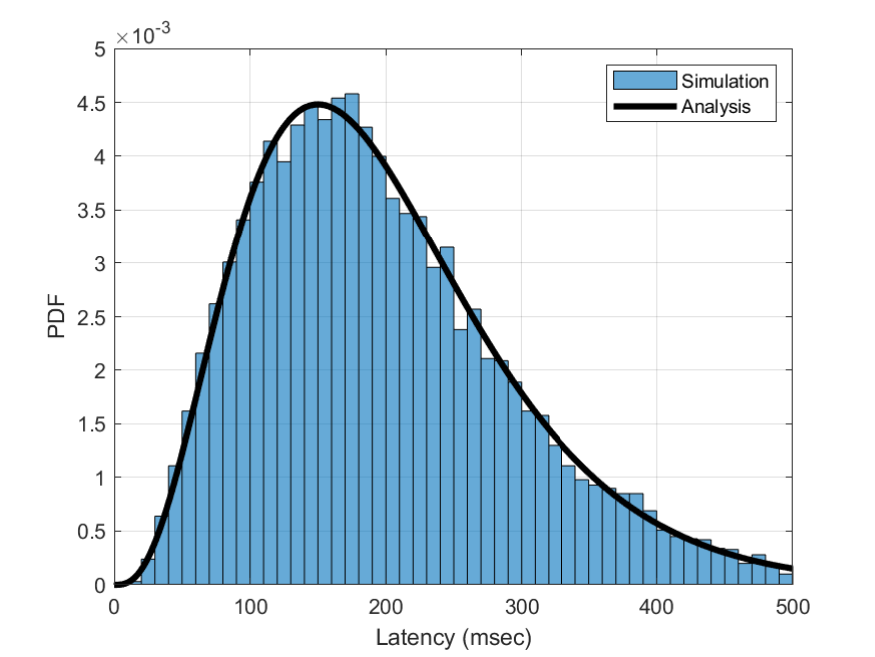}
\caption{Confirmation of Theorem \ref{theorem_gamma_sum} with simulation results: PDF of the length of time taken to reach a full SSA (The PDF shows an example of $\text{Gamma}\left(4,1/50\right)$.)}
\label{fig_PDF}
\end{figure}

%\begin{figure}
%\centering
%\includegraphics[width=\linewidth]{figs/fig_CDF_lambda}
%\caption{CDF of the length of time taken to reach a full SSA: Comparison to the latency requirements for representative ITS safety applications}
%\label{fig_CDF_lambda}
%\end{figure}

\begin{figure}
\centering
\begin{subfigure}[b]{\linewidth}
\centering
\includegraphics[width=\linewidth]{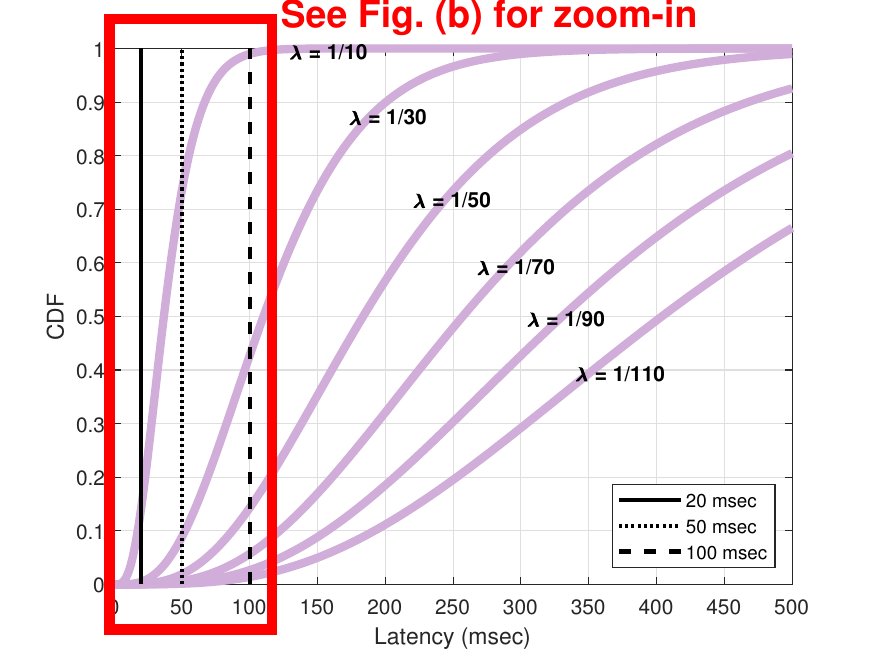}
\caption{Latency within $[0, 500]$ msec}
\label{fig_CDF_all}
\end{subfigure}
\begin{subfigure}[b]{\linewidth}
\centering
\includegraphics[width=\linewidth]{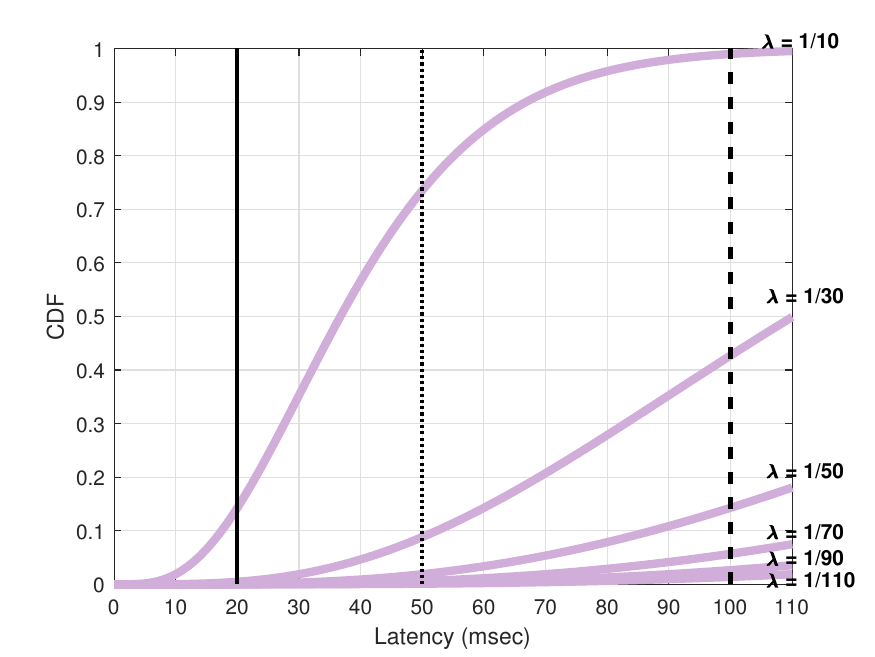}
\caption{Latency zoomed into within $[0, 110]$ msec}
\label{fig_CDF_zoomin}
\end{subfigure}
\caption{CDF of the length of time taken to reach a full SSA: Comparison to the latency requirements for representative ITS safety applications}
\label{fig_CDF}
\end{figure}

\subsection{Confirmation of Theorem \ref{theorem_gamma_sum}}\label{sec_results_theorem}
The first part of simulation is focused on confirming the analysis that in Section \ref{sec_analysis}, which led to Theorem \ref{theorem_gamma_sum}. We start from a scenario where vehicles need to share their SA in a distributed manner among themselves and eventually accomplish a complete SSA.

Recalling from Fig. \ref{fig_model}, one should observe the presence of a randomly positioned obstacle (following a uniform distribution) within the system's boundaries. This obstacle symbolizes a potential hazard such as a roadblock, construction site, or car crash, requiring acknowledgment from all vehicles. Let us assume that vehicle $i$ is the initial one to detect this hazardous object within its line of sight. Immediately upon detection, vehicle $i$ initiates the dissemination of this information through V2X communications to alert other vehicles in the system. It is important to note that each vehicle can transmit and receive messages every 100 msec. Therefore, shortly after becoming aware of the obstacle, vehicle $i$ sends a message to nearby vehicle(s) after a 100-msec interval. Upon receiving this message, other vehicles replicate the process, transmitting the information to their neighboring vehicles, each step consuming another time slot. The propagation cycle concludes when the first vehicle achieves a complete SSA. It is crucial to emphasize that this scenario mirrors the sequential dissemination of knowledge along a path, reaching the last vehicle in the process.

Fig. \ref{fig_PDF} shows the distribution of latency for a distributed V2X network to reach a complete SSA, wherein the black curve indicates $\text{Gamma}\left(4, 1/50\right)$ and the histogram gives the result of simulation with randomly distributed vehicles in $\mathbb{A}^{2}$. As the figure depicts, the distribution that was found in Theorem \ref{theorem_gamma_sum} fits very precisely the result induced from simulations.

\subsection{Variation of Vehicle Density $\lambda$}\label{sec_results_lambda}
Now, with the confirmation of the analysis with computer simulations, we proceed to varying the vehicle density, $\lambda$, which forms the predominant factor determining the performance of SSA and distributed V2X networking. Fig. \ref{fig_CDF} presents the result of this variation in $\lambda$.

In Fig. \ref{fig_CDF_all}, the distribution of latency is shown in its CDF. Notice that $\lambda = \left\{1/10, 1/30, 1/50, 1/70, 1/90, 1/110\right\}$ indicates $100e3$, $33.\bar{3}e3$, $20e3$, $14.2857e3$, $11.\bar{1}e3$, and $9.0909e3$ vehicles, respectively, in $\left|\mathbb{A}^2\right| = 1 \hspace{0.02 in} \text{km} \times 1 \hspace{0.02 in} \text{km}$. These numbers look unrealistically large at first glance but they translate to $\left\{316.23, 182.5741, 141.4215, 119.5228, 105.4092, 95.3462 \right\}$ vehicles within a 1-km line segment, respectively, which represent quite a breadth of traffic densities from extremely heavy traffic (i.e., $\lambda = 1/10$) to fairly light traffic (i.e., $\lambda = 1/110$). With that said, from Fig. \ref{fig_CDF_all}, one can find a clear tendency from the comparisons: a higher vehicle density yields a larger portion of a V2X network to the ability of supporting a safety application. This pattern can be rationalized as follows: a denser network leads to an easier environment to find the next vehicle to hand in the fragment of knowledge, which, in turn, makes it easier to complete a full SSA within a shorter time.

In Fig. \ref{fig_CDF_zoomin}, we zoom into the range of $[0, 110]$ msec of time in order to make direct comparison to latency requirements of some representative applications for traffic safety and autonomous driving \cite{yonsei}. This comparison enables us to scrutinize the feasibility of the SSA in such critical ITS applications. The reference \cite{yonsei} indicates that use cases including forward collision warning, emergency stop, and cooperative collision avoidance require 100 msec of latency. There are other use cases that require shorter latencies: 50 msec for see-through, 20 msec for pre-crash sensing warning, and 10 msec for automated overtake and high-density platooning \cite{sae2735}. With these requirements considered, we evaluate how much of a V2X network can achieve a full SSA within a latency that is short enough to support the applications.

The overall implication of Fig. \ref{fig_CDF} is that it takes shorter to complete a SSA as the network is more highly populated. The rationale is straightforward: as a network is populated with a larger number of vehicles, more neighboring vehicles can be found, which yields a higher chance for the pieces of SA to be combined into a full SSA.

%%%%%%%%%%%%%%%%%%%%%%%%%%%%%%%%%%%%%%%%%%%%%%%%%%%%%%%%%%%%%%%%%%%%%%%%%%%%%%%%%%%%%%%%%%%%%%%%%%%%%%%%%%%%%%%%%%%%
\section{Concluding Remarks}\label{sec_conclusions}

This paper provided an accurate analysis framework that calculates the latency for completion of a SSA among a group of connected vehicles. It discovered that the latency was characterized as a Gamma random variable. Our theoretical analysis was confirmed by computer simulations with randomly distributed vehicles. The simulation was extended to variation of vehicle density, which was compared to the latency requirements defined in the SAE J2735---viz., 20, 50, and 100 msec. The paper concluded that a higher vehicle density leads to a shorter SSA latency, owing to a higher chance of being able to find a neighboring vehicle and thus a faster completion of combined SSA.

Future work can be directed to increasing the realism of use cases. We will build on this paper's system model to add applications into consideration. Recall that this paper has found that a higher vehicle density would benefit in obtaining a SSA. This cooperative intelligence will increase the accuracy of reinforcement learning \cite{vtc21} that will be adopted as a key facilitator technique for autonomous driving. Blockchain \cite{iceic22}-\cite{iv24} is another technique that has potential to be appleid to ITS as a method to promote the trust among vehicles.

%%%%%%%%%%%%%%%%%%%%%%%%%%%%%%%%%%%%%%%%%%%%%%%%%%%%%%%%%%%%%%%%%%%%%%%%%%%%%%%%%%%%%%%%%%%%%%%%%%%%%%%%%%%%%%%%%%%%

\end{document}